\documentstyle[11pt]{article}
\begin{document}
\title{Semiclassical description of Heisenberg models\\
           via spin--coherent states}
\author{John Schliemann and Franz G. Mertens\\
{\it Physikalisches Institut, Universit{\"a}t Bayreuth, D-95440 Bayreuth,
Germany}}
\date{November 1997}
\maketitle
\begin{abstract} 
We use spin--coherent states as a time-dependent variational ansatz for
a semiclassical description of a large family of Heisenberg models. In
addition to common approaches we also evaluate the square variance
$\langle{\cal H}^2\rangle-\langle{\cal H}\rangle^2$ of the Hamiltonian
in terms of coherent states. This quantity turns out to have a natural
interpretation with respect to time-dependent solutions of the equations
of motion and allows for an estimate of quantum fluctuations in a 
semiclassical regime. The general results are applied to solitons,
instantons and vortices in several one- and two--dimensional models.
\end{abstract}

%%%%%%%%%%%%%%%%%%%%%%%%%%%%%%%%%%%%%%%%%%%%%%%%%%%%%%%%%%%%%%%%%%%%%%

\section{Introduction}
There has been a lot of study on classical spin systems in the last decades,
commonly using a continuum description, for a review see Kosevich {\it et al.}
1990. Some one--dimensional models have been identified as integrable and 
treated by means of the inverse scattering method (Takhtajan 1977,
Fogedby 1980, Sklyanin 1979). A common feature of such 
nonlinear evolution equations is the existence of localized solutions 
like solitary waves and, in the integrable case, mathematical solitons.\\
In quantum spin systems analytical treatments appear to be
restricted to one--dimensional models. For the spin--$\frac{1}{2}$--Heisenberg 
chain of $N$ lattice sites Bethe (1931) managed to
reduce the diagonalisation of the Hamiltonian acting on a $2^N$--dimensional 
Hilbert space to the problem of solving $N$ coupled nonlinear algebraic 
equations, 
which are well understood in the thermodynamic limit. By this Bethe ansatz and
its algebraic formulation many properties of the model such as thermodynamic
quantities could be extracted. Nevertheless,
the eigenstates of the Hamiltonian themselves are obtained only rather 
implicitly and are naturally translationally symmetric, i~.e. non localized.
Thus, a quantum analogon to a classical time--dependent soliton, which
should not be an eigenstate of energy but a superposition of them, has 
not been found yet.\\ 
So it is desirable to give a description of solitons in the semiclassical
regime, i.e. for large but finite values of the spin length.
Considerable work in this direction has been done using different kinds
of bosonization and tensor products of coherent oscillator states
as variational ansatz, see e.~g. De Azevedo {\it et al.} 1982, Ferrer 
and Pozo 1988, Kapor {\it et al.} 1990. Disadvantages of these methods
are difficulties in treating infinite series of nontrivial operator
products and, more fundamental, the problem of mapping the finite
dimensional Hilbert space of a spin onto an infinite dimensional bosonic 
space.\\
An alternative to such approaches are spin--coherent states as introduced
by Radcliffe (1971). Balakrishnan, Bishop (1985, 89) and Balakrishnan
{\it et al.} (1990)
used these objects for treating ferromagnetic Heisenberg chains and
evaluated quantum corrections to classical soliton dispersion laws.
In detail, their way of proceeding has been subject to some criticism
by Haldane (1986), but the general findings concerning soliton
stability with respect to quantum fluctuation seem to be valid and
will be confirmed in the present work. Moreover, Frahm and Holyst (1989)
suggested an extension of spin--coherent states introducing
a squeezing similar to bosonic states. Unfortunately, this approach leads
for generic models to very complicated equations of motion for the 
classical angular variables and the additional squeeze parameter, that cannot
be solved exactly for nontrivial cases.\\
In this paper we follow the above authors and use (unsqueezed) spin--coherent
states as variational ansatz for a large family of Heisenberg models. In
addition to the quantities considered in the above
references we also calculate the quantum mechanical variance 
of the energy. This quantity can either be used as a test of the validity
of the approach; on the other hand, the general expression obtained has
a natural physical interpretation concerning energy fluctuations in
time--dependent spin structures. Finally we apply our results to several
prominent solitary solutions of spin models.

%%%%%%%%%%%%%%%%%%%%%%%%%%%%%%%%%%%%%%%%%%%%%%%%%%%%%%%%%%%%%%%%%%%%%%

\section{Coherent states}
In the Hilbert space of a spin of length $S$ we define a spin--coherent 
state $|S;\vartheta,\varphi\rangle$ by the equation
\begin{equation}
\vec s_{\vartheta,\varphi}\cdot\hat{\vec S}\,|S;\vartheta,\varphi\rangle
=\hbar S\,|S;\vartheta,\varphi\rangle
\label{def1}
\end{equation}
for the direction $\vec s_{\vartheta,\varphi}=
(\sin\vartheta\cos\varphi,\sin\vartheta\sin\varphi,\cos\vartheta)$.
In the usual basis of eigenstates of $\hat S^z$ 
($\hat S^z \,|m\rangle=\hbar m\,|m\rangle$) these states can be
expressed as
\begin{equation} 
|S;\vartheta,\varphi\rangle=U(\vartheta,\varphi)\,|S\rangle=
\sum_{n=0}^{2S}{2S\choose n}^{\frac{1}{2}}
\left(\cos\left(\frac{\vartheta}{2}\right)\right)^{2S-n}
\left(\sin\left(\frac{\vartheta}{2}\right)\right)^{n}
e^{\imath\varphi(n-S)}\,|S-n\rangle
\end{equation}
with 
\begin{equation}
U(\vartheta,\varphi)=\exp\left(-\frac{\imath}{\hbar}\varphi\hat S^z\right)
\exp\left(-\frac{\imath}{\hbar}\vartheta\hat S^y\right)\,.
\label{defU}
\end{equation}
Clearly, we have $\langle S;\vartheta,\varphi|\hat{\vec S}\,
|S;\vartheta,\varphi\rangle=\hbar S\vec s_{\vartheta,\varphi}$,
but in the expectation values of higher products of spin components 
contributions of different order in the spin length $S$ arise.
A list of diagonal elements useful for the following is given appendix
\ref{app1}, where we also demonstrate the classical limit of
such expressions.\\
The spin--coherent states have the minimum uncertaincy product
\begin{equation}
\Delta(\vec e_{1}\cdot\hat{\vec S}\,)\,\Delta(\vec e_{2}\cdot\hat{\vec S}\,)=
\frac{\hbar}{2}\Delta(\vec e_{3}\cdot\hat{\vec S}\,)
\end{equation}
with $\Delta(\hat O)$ being the variance of an operator $\hat O$ and
$\vec e_{1}$,$\vec e_{2}$,$\vec e_{3}$ an orthonormal system.
Furthermore, they fullfill the (over-) completeness relation
\begin{equation}
\frac{2S+1}{4\pi}\,\int_0^{2\pi}d\varphi\int_0^{\pi}d\vartheta\sin\vartheta
|S;\vartheta,\varphi\rangle\langle S;\vartheta,\varphi|=\bf 1
\end{equation}
Nevertheless, it should be mentioned that for an arbitrary linear 
combination of coherent states a direction $\vec s_{\vartheta,\varphi}$
solving eq. (\ref{def1}) cannot be found. Therefore the spin--coherent states
do not form a subspace of the Hilbert space, but a submanifold
diffeomorphic to the two--dimensional unit sphere. Only in the case 
$S=\frac{1}{2}$ every (normalized) state vector can be identified as a
coherent state.

%%%%%%%%%%%%%%%%%%%%%%%%%%%%%%%%%%%%%%%%%%%%%%%%%%%%%%%%%%%%%%%%%%%%%

\section{Time--dependent variational method}

Let us consider the following class of one--dimensional spin models

\begin{eqnarray}
{\cal H}=\sum_{n}\sum_{i=1}^{3}\left[\alpha_{i}\hat S_{n}^{i}\hat S_{n+1}^{i}
+\beta_{i}\hat S_{n}^{i}\hat S_{n}^{i}+\gamma_{i}\hat S_{n}^{i}\right]\,.
\label{defmod1}
\end{eqnarray}
The index $i$ runs over the three spatial directions $x,y,z$. Each lattice site
labelled by $n$ carries a spin of uniform length $S$. The Hamiltonian
includes an anisotropic exchange coupling between nearest neighbors, a local
anisotropy and a magnetic field. The parameters
$\alpha_{i}$, $\beta_{i}$, $\gamma_{i}$ may be chosen arbitrarily, in
particular with respect to their sign.\\
We now employ the direct product of coherent states 
\begin{equation}
|\psi(t)\rangle=\bigotimes_{n}|S;\vartheta_n,\varphi_{n}\rangle
\label{defvar}
\end{equation}
as a time--dependent variational ansatz, i.e. we assume the time evolution
of the state $|\psi(t)\rangle$ under the above Hamiltonian to be given
in terms of time--dependent functions $\vartheta_{n}(t)$, $\varphi_{n}(t)$.\\
With $\langle\cdot\rangle$ denoting an expectation value within 
(\ref{defvar}), 
$\vec s_{n}=(\sin\vartheta_{n}\cos\varphi_{n},\sin\vartheta_{n}
\sin\varphi_{n},\cos\vartheta_{n})$ and cartesian directions $\vec e^{\,i}$, 
we have for $\cal H$
\begin{equation}
\langle{\cal H}\rangle=(\hbar S)^2\sum_{n}\sum_{i=1}^{3}
\left[\alpha_{i}\left(\vec s_{n}\cdot\vec e^{\,i}\right)
\left(\vec s_{n+1}\cdot\vec e^{\,i}\right)+\left(1-\frac{1}{2S}\right)
\beta_{i}\left(\vec s_{n}\cdot\vec e^{\,i}\right)^2
+\frac{\gamma_{i}}{\hbar S}\left(\vec s_{n}\cdot\vec e^{\,i}\right)\right]\,,
\label{energy1}
\end{equation}
and from the Heisenberg equations of motion 
$\frac{d}{dt}\hat{\vec S_{n}}=-\frac{\imath}{\hbar}
[\hat{\vec S_{n}},{\cal H}]$ :
\begin{eqnarray}
\frac{d}{dt}\vec s_{n}&=&(\hbar S){\sum_{i=1}^{3}}
\biggl[\alpha_{i}\left(\vec s_{n}\times\vec e^{\,i}\right)
\left(\left(\vec s_{n-1}+\vec s_{n+1}\right)\cdot\vec e^{\,i}\right)
+2\left(1-\frac{1}{2S}\right)\beta_{i}
\left(\vec s_{n}\times\vec e^{\,i}\right)
\left(\vec s_{n}\cdot\vec e^{\,i}\right)\nonumber\\
 & & \hspace{3cm}+\frac{\gamma_i}{\hbar S}\left({\vec s_{n}}
\times{\vec e^{\,i}}\right)\biggr]
\label{eom1}
\end{eqnarray}
The quantity $(\hbar S)$ becomes the classical spin length in the limit
$S\to\infty$, $\hbar\to 0$, $\hbar S={\rm const.}$
Eqs.~(\ref{energy1}), (\ref{eom1}) are identical to the classical energy and 
the Landau--Lifshitz--equation 
up to a renormalisation of the local anisotropy by a factor 
$(1-\frac{1}{2S})$, reflecting the fact that this term does not 
contribute to the dynamics for $S=\frac{1}{2}$. Therefore the variational 
ansatz provides semiclassical corrections to the equation of motion and
reproduces the correct classical limit. This property of
spin--coherent states has been found by several authors before with respect 
to particular spin models, see e.~g. the references given in the introduction,
and is proved in appendix \ref{app2} for
an arbitrary Hamiltonian.\\
Next we examine the square variance of the Hamiltonian, i.~e.
$\langle{\cal H}^2\rangle-\langle{\cal H}\rangle^2$.
This quantity is nonzero only in the quantum case and (as well as 
$\langle{\cal H}\rangle$) strictly an invariant of the system, whatever
the exact quantum mechanical time evolution of the state (\ref{defvar}) 
will be. After extensive algebra one ends up with the following 
expression:
\begin{eqnarray}
 & & \langle{\cal H}^2\rangle-\langle{\cal H}\rangle^2=\nonumber\\ 
 & & (\hbar S)^4\sum_{n}\Biggl[\frac{1}{2S}\Biggl(\sum_{i=1}^{3}\Big[
\alpha_{i}\left(\vec s_{n}\times\vec e^{\,i}\right)
\left(\left(\vec s_{n-1}+\vec s_{n+1}\right)\cdot\vec e^{\,i}\right)
\nonumber\\
 & & \hspace{2cm}+2\left(1-\frac{1}{2S}\right)\beta_{i}
\left(\vec s_{n}\times\vec e^{\,i}\right)
\left(\vec s_{n}\cdot\vec e^{\,i}\right)
+\frac{\gamma_i}{\hbar S}\left({\vec s_{n}}
\times{\vec e^{\,i}}\right)\Big]\Biggr)^2\nonumber\\
 & & \hspace{1.5cm}+\frac{1}{8S^2}\Biggl(
\sum_{i,j=1}^{3}\Big[\alpha_{i}\alpha_{j}
\Bigl(\left(\left(\vec s_{n}\times\vec e^{\,i}\right)\cdot
\left(\vec s_{n}\times\vec e^{\,j}\right)\right)\nonumber\\
 & & \hspace{5.3cm}\left(\left(\vec s_{n-1}\times\vec e^{\,i}\right)\cdot
\left(\vec s_{n-1}\times\vec e^{\,j}\right)
+\left(\vec s_{n+1}\times\vec e^{\,i}\right)\cdot
\left(\vec s_{n+1}\times\vec e^{\,j}\right)\right)\nonumber\\
 & & \hspace{4.5cm}-\left(\vec s_{n}\cdot
\left(\vec e^{\,i}\times\vec e^{\,j}\right)\right)
\left(\left(\vec s_{n-1}+\vec s_{n+1}\right)\cdot
\left(\vec e^{\,i}\times\vec e^{\,j}\right)\right)
\Bigr)\Big]\nonumber\\
 & & \hspace{2.2cm}+2\sum_{i,j=1}^{3}\Big[
\left(1-\frac{1}{2S}\right)\beta_{i}\beta_{j}\left(
\left(\vec s_{n}\times\vec e^{\,i}\right)^{2}
\left(\vec s_{n}\times\vec e^{\,j}\right)^{2}
-\left(\vec s_{n}\cdot
\left(\vec e^{\,i}\times\vec e^{\,j}\right)\right)^{2}\right)\Big]\Biggr)
\Biggr]
\nonumber\\
\label{var1}
\end{eqnarray} 
The square variance of the energy consists essentially of two contributions
being of order $\frac{1}{S}$ and $\frac{1}{S^2}$ respectively. In the
above summation over the lattice sites the squared expression in the 
term of order $\frac{1}{S}$ can be recognized as the r.h.s. of
eq.~(\ref{eom1}). Thus, we have
\begin{equation}
\langle{\cal H}^2\rangle-\langle{\cal H}\rangle^2=(\hbar S)^2
\sum_{n}\frac{1}{2S}\left(\frac{d}{dt}\vec s_{n}\right)^2
\,+\,O\left(\frac{1}{S^2}\right)\,.
\label{var2}
\end{equation}
Within our variational approach the leading order of the quantum fluctuations
of the energy is purely due to the time--dependence of the state vector. On
the other hand, for a quantum state which has a nontrivial time evolution
and is consequently not an eigenstate of the Hamiltonian, the energy
must definetely have a finite uncertaincy. Following this observation the
leading term in (\ref{var2}) is certainly not an artifact of the
ansatz (\ref{defvar}), but it is a physically relevant expression for 
the energy fluctuations in time--dependent semiclassical spin structures. 
So we have found good evidence
that the ansatz of spin--coherent states does not only reproduce
the classical limit but is still meaningfull for a semiclassical description.
The contributions of order $\frac{1}{S^2}$ in (\ref{var1}) cannot be 
interpretated in a general way and should be studied 
in the context of particular models.
We will see that these terms can often be considered as a criterion for 
the validity of the variational method.\\
For brevity we have concentrated in (\ref{defmod1}) on an one--dimensional
model. For higher dimensions one simply has to infer the appropriate
number of neighbors of each lattice site in the summations. The result can
always be written in the form (\ref{var2}). 
Moreover, it is also straightforeward to see that the result (\ref{var2}) is
still valid in the case of exchange couplings of longer range.\\
Thus, our above findings apply to a large variety of 
(ferromagnetic or antiferromagnetic) spin models in arbitrary dimension.
This will be illustrated in the next section with respect to several
one-- and two--dimensional ferromagnetic models.

%%%%%%%%%%%%%%%%%%%%%%%%%%%%%%%%%%%%%%%%%%%%%%%%%%%%%%%%%%%%%%%%%%%%%%

\section{Application to solitons, instantons and vortices}
We now calculate the quantum mechanical energy uncertaincy for
solitary solutions in different Heisenberg models. We denote the leading
order in $\frac{1}{S}$ by $\Omega_{1}$ (cf.~eq.~(\ref{var2})), the remaining
contributions by $\Omega_{2}$, i.e.
$\langle{\cal H}^2\rangle-\langle{\cal H}\rangle^2=\Omega_{1}+\Omega_{2}$.\\\\
\underline{$(i)$}\,  A ferromagnetic Heisenberg chain with isotropic
exchange coupling and an external field is given by
\begin{equation}
{\cal H}_{1}=-J\sum_{n}\left[\hat{\vec S}_{n}\hat{\vec S}_{n+1}
+\hat{\vec S}_{n}\vec B\right]\,,
\label{defmod2}
\end{equation}
with $J>0$. Eqs.~(\ref{energy1})--(\ref{var1}) read:
\begin{eqnarray}
\langle{\cal H}_{1}\rangle&=&-{(\hbar S)^2}J
\sum_{n}\left[\vec s_{n}\vec s_{n+1}
+\vec s_{n}\frac{\vec B}{\hbar S}\right]\\
\frac{d}{dt}\vec s_{n}&=&(\hbar S)J\left(\vec s_{n}
\times\left(\vec s_{n-1}+\vec s_{n+1}\right)+\vec s_{n}\times
\frac{\vec B}{\hbar S}\right)
\label{eom2}\\
\langle{\cal H}_{1}^2\rangle-\langle{\cal H}_{1}\rangle^2&=&(\hbar S)^{4}J^2
\sum_{n}\Biggl[\frac{1}{2S}\left(\vec s_{n}
\times\left(\vec s_{n-1}+\vec s_{n+1}\right)+\vec s_{n}\times
\frac{\vec B}{\hbar S}\right)^2\nonumber\\
& & \hspace{2cm}+\frac{1}{8S^2}\left(\left(1-\vec s_{n}\vec s_{n-1}\right)^{2}
+\left(1-\vec s_{n}\vec s_{n+1}\right)^{2}\right)\Biggl]
\label{var3}
\end{eqnarray} 
We choose appropriate units with $J=1$, $\hbar S=1$ and
the lattice spacing $a=1$ and calculate the expectation value of 
(\ref{defmod2}) in the usual continuum approximation for an infinite system
\begin{equation}
\langle{\cal H}_{1}\rangle=\int d\xi\left(\frac{1}{2}\left(
\frac{\left(\partial_{\xi}p\right)^2}{1-p^2}+\left(1-p^2\right)
\left(\partial_{\xi}q\right)^2\right)-Bp\right)\,,
\label{energy2}
\end{equation}
where we have put the magnetic field in $z$ direction,
substracted the ground state energy
and introduced the canonical conjugate fields $p=\cos\vartheta$,
$q=\varphi$; $\xi$ denotes the spatial variable in chain direction.
The equations of motion (\ref{eom2}) read in continuum approximation:
\begin{eqnarray}
\partial_{t}q&=&-\frac{1}{1-p^2}\partial_{\xi}^{2}p
-\frac{p}{(1-p^2)^2}(\partial_{\xi}p)^2-p(\partial_{\xi}q)^2-B\\
\partial_{t}p&=&(1-p^2)\partial_{\xi}^{2}q-2p(\partial_{\xi}p)\partial_{\xi}q
\end{eqnarray}
These equations have well--known soliton solutions (Tjon and Wright 1977):
\begin{eqnarray}
p(\xi,t)&=&1-\frac{2}{\left(B+\omega\right)\Gamma^2}
{\rm sech}^{2}\left(\frac{\xi-vt-\xi_0}{\Gamma}\right)\\
q(\xi,t)&=&q_{0}+\omega t+\frac{v}{2}\left(\xi-vt-\xi_0\right)
+\tan^{-1}\left(\frac{2}{v\Gamma}\tanh\left(\frac{\xi-vt-\xi_0}{\Gamma}
\right)\right)
\end{eqnarray}
with the soliton width
\begin{eqnarray}
\Gamma=\frac{1}{\sqrt{B+\omega-\frac{v^2}{4}}}>0
\label{constr}
\end{eqnarray}
and $\xi_0$, $q_0$ being constants. The above solution is a pulse soliton
with velocity $v$ and an internal frequency $\omega$ constrained by
(\ref{constr}). Its energy is calculated from (\ref{energy2}):
\begin{equation}
E=\frac{4}{\Gamma}+\frac{B}{B+\omega}
\frac{4}{\Gamma}
\end{equation}
The square variance
$\langle{\cal H}_{1}^2\rangle-\langle{\cal H}_{1}\rangle^2=\Omega_{1}
+\Omega_{2}$ can be obtained from the continuum version of (\ref{var3}):
\begin{eqnarray}
\Omega_{1}&=&\frac{1}{S}\left(\frac{1}{\Gamma}\left(8\left(2B+3\omega\right)
+\frac{4\omega^2}{B+\omega}\right)-\frac{1}{\Gamma^3}
\left(16+\frac{8\omega^2}{3(B+\omega)^2}
+\frac{8\omega}{B+\omega}
\right)\right)\,,\\
\Omega_{2}&=&\frac{1}{S^2}\frac{4}{3\Gamma^3}\,,
\end{eqnarray}
where (\ref{constr}) has been used. As explained in the previous section
the quantity $\Omega_{1}$ can be interpreted as a natural property of a 
time--dependent spin state in the semiclassical regime. As a criterion 
for the importance of 
quantum effects absent from our ansatz (\ref{defvar}) we compare $\Omega_{2}$
with the soliton energy, i.~e.
\begin{equation}
\frac{\sqrt{\Omega_{2}}}{E}
=\frac{1}{S\sqrt{\Gamma}}\frac{B+\omega}{2\sqrt{3}(2B+\omega)}
\end{equation}
We see that higher order quantum corrections to our variational approach are
neglegible for solitons with large width and consequently small energy.
If this condition is not fullfilled the spin length $S$ becomes significant,
in agreement with the work of Balakrishnan and Bishop mentioned in the 
introduction. \\
The continuum approximation is a good description of the system if the
classical fields $p$, $q$ vary only weakly on the typical length scale.
This means that the soliton width should be significantly larger than the
lattice spacing $a=1$, e.g. $\Gamma>10$. Thus, continuum and semiclassical 
approximation work well in a consistent area of the soliton parameters.\\\\
\underline{$(ii)$}\, Next we consider a ferromagnetic Heisenberg chain
with a biaxial local anisotropy:
\begin{equation}
{\cal H}_{2}=-J\sum_{n}\left[\hat{\vec S}_{n}\hat{\vec S}_{n+1}
+\tau_{x}\hat S_{n}^{x}\hat S_{n}^{x}+\tau_{z}\hat S_{n}^{z}\hat S_{n}^{z}
\right]
\label{defmod3}
\end{equation}
\pagebreak
The square variance of the energy is given by
\begin{eqnarray}
 & & \langle{\cal H}_{2}^2\rangle-\langle{\cal H}_{2}\rangle^2=\nonumber\\
 & & (\hbar S)^{4}J^2
\sum_{n}\Biggl[\frac{1}{2S}\Biggl(\vec s_{n}
\times\left(\vec s_{n-1}+\vec s_{n+1}\right)\nonumber\\
 & & \hspace{3cm}
+2\left(1-\frac{1}{2S}\right)\left(
\tau_{x}\left(\vec s_{n}\times\vec e^{\,x}\right)
\left(\vec s_{n}\cdot\vec e^{\,x}\right)
+\tau_{z}\left(\vec s_{n}\times\vec e^{\,z}\right)
\left(\vec s_{n}\cdot\vec e^{\,z}\right)\right)\Biggr)^2\nonumber\\
 & & \hspace{2cm}+\frac{1}{8S^2}
\Biggl(\left(1-\vec s_{n}\vec s_{n-1}\right)^{2}
+\left(1-\vec s_{n}\vec s_{n+1}\right)^{2}\nonumber\\
 & & \hspace{3cm}
+2\left(1-\frac{1}{2S}\right)\Bigl(\tau_{x}^{2}
\left(\vec s_{n}\times\vec e^{\,x}\right)^{4}
+\tau_{z}^{2}\left(\vec s_{n}\times\vec e^{\,z}\right)^{4}\nonumber\\
 & & \hspace{4.5cm}+2\tau_{x}\tau_{z}
\left(\left(\vec s_{n}\cdot\vec e^{\,x}\right)
\left(\vec s_{n}\cdot\vec e^{\,z}\right)
-\left(\vec s_{n}\cdot\left(\vec e^{\,x}\times\vec e^{\,z}\right)\right)^{2}
\right)\Bigr)\Biggr)\Biggl]\nonumber\\
\label{var4}
\end{eqnarray}
Proceeding as above one can derive the following solutions of the 
continuum model (see e.~g. Kosevich {\it et al.} 1990):
\begin{equation}
p(\xi,t)={\rm sign}(v)\tanh\left(\frac{\xi-vt+\xi_0}{\Gamma}\right)
\quad,\quad q(\xi,t)=q_0
\label{kink1}
\end{equation}
with
\begin{equation}
\Gamma=\frac{1}{\sqrt{2(1-\frac{1}{2S})(\tau_{z}-\tau_{x}\cos^{2}q_{0})}}
\quad,\quad
v^{2}=\frac{\left(\tau_{x}\sin(2q_{0})\right)^{2}(1-\frac{1}{2S})}
{2\left(\tau_{z}-\tau_{x}\cos^{2}q_{0}\right)}\,,
\end{equation}
$\xi_{0}$, $q_{0}$ being constants and $\tau_{z}-\tau_{x}\cos^{2}q_{0}>0$.
The eqs.~(\ref{kink1}) describe moving kink solitons parametrized by $q_0$ 
with energy $E=\frac{2}{\Gamma}$ above the ground state.
For finite $\tau_{x}$ the velocity $v$ vanishes for 
$q_{0}\in\{0,\frac{\pi}{2}\}$ and takes its maximum 
$v_{\rm max}^{2}=|\tau_{x}\cos(2q^{\ast})|$ at $q_{0}=q^{\ast}$ with
\begin{equation}
\cos(2q^{\ast})=2\frac{\tau_z}{\tau_x}-1
-{\rm sign}(\frac{\tau_z}{\tau_x})\sqrt{(2\frac{\tau_z}{\tau_x}-1)^2-1}
\end{equation}
Inserting eqs.~(\ref{kink1}) into the continuum version of (\ref{var4}) leads
to a divergence in $\Omega_{2}$ arising from the term proportional to 
$\tau_{x}^{2}$. This contribution is almost uniform over the whole system
except for an area around the center of the kink. Moreover, the same
divergence occurs if we evaluate $\Omega_{2}$ for the classical ground 
state of the model (\ref{defmod3}). The explanation for this behaviour is 
that, different from
the previous model, the quantum mechanical ground state of (\ref{defmod3})
is not described exactly by our ansatz (\ref{defvar}). Thus, the divergent
contribution in order $\frac{1}{S^2}$ is clearly an artifact of our
variational approach due to a more complicated structure of the 
quantum mechanical ground state, which does not correspond trivially
to the classical ground state solution. Note that a similar infinite term
in $\Omega_{2}$ arises for the classical ground state of an antiferromagnetic 
Heisenberg model, e.~g. (\ref{defmod2}) with $J<0$.
Here the divergence in order $\frac{1}{S^2}$ is due to the contribution
from the isotropic exchange in the Hamiltonian.\\
Since we are interested in
physically valid quantum fluctuations in time--dependent excitations
within a semiclassical approach, we renormalize 
$\langle{\cal H}_{2}^2\rangle-\langle{\cal H}_{2}\rangle^2$ by neglecting
$\Omega_{2}$. The remaining contribution is calculated easily giving
\begin{equation}
\Omega_{1}=\frac{1}{S}\frac{v^2}{\Gamma}=\frac{1}{S}
\frac{\left(\tau_{x}\sin(2q_{0})\right)^2\left(1-\frac{1}{2S}\right)
^{\frac{3}{2}}}{\left(2\left(\tau_{z}-\tau_{x}\cos^{2}q_{0}\right)\right)
^{\frac{1}{2}}}\,\,.
\end{equation}  
This leading order of energy variance vanishes of course
in the static case $q_{0}\in\{0,\frac{\pi}{2}\}$ and takes its maximum
at $q_{0}=q^{\ast\ast}$ where
\begin{equation}
\cos(2q^{\ast\ast})=2\left(2\frac{\tau_z}{\tau_x}-1\right)
-2\,{\rm sign}(\frac{\tau_z}{\tau_x})
\sqrt{(2\frac{\tau_z}{\tau_x}-1)^2-\frac{3}{4}}
\end{equation}
Obviously, $q^{\ast}$ and $q^{\ast\ast}$ do not coincide.\\
In the case $\tau_{x}=0$ the kink solitons become static, and the exact
quantum ground state is again the trivial ferromagnetic state. Thus,
$\Omega_{1}=0$ and in $\Omega_{2}$ no unphysical divergency arises:
\begin{equation}
\Omega_{2}=\frac{1}{S^2}\frac{1}{3\sqrt{2}}\sqrt{\left(1-\frac{1}{2S}\right)}
\tau_{z}^{\frac{3}{2}}\left(1+4\left(1-\frac{1}{2S}\right)\right)
\end{equation}
Comparing $\sqrt{\Omega_{2}}$ with the energy of the kink excitation, we
again conclude that the semiclassical description should be valid 
for solitons with large width and low energy.\\\\
\underline{$(iii)$}\, The isotropic Heisenberg ferromagnet on a 
two--dimensional infinite square lattice is given by
\begin{equation}
{\cal H}_{3}=-J\sum_{m,n}\hat{\vec S}_{m,n}\left(\hat{\vec S}_{m+1,n}
+\hat{\vec S}_{m,n+1}\right)
\label{defmod4}
\end{equation}
In appropriate units we find in the continuum approximation
\begin{equation}
\langle{\cal H}_{3}\rangle=\frac{1}{2}\int d^{2}x\left(
\frac{\left(\nabla p\right)^2}{1-p^2}+\left(1-p^2\right)
\left(\nabla q\right)^2\right)
\label{energy3}
\end{equation}
Eq.~(\ref{energy3}) defines a two--dimensional conformal invariant field
theory that has been investigated by Belavin and Polyakov (1975).
The fields $p$, $q$ parametrize points on the unit sphere $S^{2}$. 
Compactifying the two--dimensional plane by an infinite point we can
consider the solutions of the Hamiltonian (\ref{energy3}) as mappings
$S^{2}\rightarrow S^{2}$. These mappings can be classified in 
homotopical classes characterized by the degree of mapping $d$ with
\begin{equation}
|d|=\frac{1}{8\pi}\int d^{2}x\,\varepsilon_{\alpha\beta\gamma}\,
\varepsilon_{\mu\nu}s^{\alpha}\frac{\partial s^{\beta}}{\partial x_{\mu}}
\frac{\partial s^{\gamma}}{\partial x_{\nu}}
=\frac{1}{4\pi}\int d\varphi(\vec x)\,d(\cos\vartheta(\vec x))\,,
\end{equation}
which is the number of times that the sphere is covered in the course
of mapping. As shown by the above authors it holds
\begin{equation}
\langle{\cal H}_{3}\rangle=4\int d^{2}z\frac{1}{\left(1+|w|^{2}\right)^2}
\left(\frac{\partial w}{\partial z}\frac{\partial\bar w}{\partial\bar z}
+\frac{\partial\bar w}{\partial z}\frac{\partial w}{\partial\bar z}\right)
\geq 4\pi |d|
\label{deg}
\end{equation}
with $w(z,\bar z)=\cot\frac{\vartheta}{2}e^{\imath\varphi}$ and
$z=x_{1}+\imath x_{2}$. The minimum
in (\ref{deg}) is realized by instanton solutions that are given by
arbitrary meromorphic functions $w(z)$ :
\begin{equation}
w(z)=\prod_{i}\left(\frac{z-a_{i}}{R}\right)^{m_{i}}
\cdot\prod_{i}\left(\frac{R}{z-b_{i}}\right)^{n_{i}}
\end{equation}
with $m_{i}, n_{i}>0$ and $a_{i}\neq b_{j}$ for all $i,j$ and the degree
given by $d=\max\{\sum_{i}m_{i},\sum_{i}n_{i}\}$.  
The real scale parameter $R$ does not influence the energy of these static
excitations due to the conformal invariance of the model. \\
Let us now consider a single instanton simply given by 
$w=\left(\frac{z}{R}\right)^{\nu}$, $d=|\nu|$, or
\begin{equation}
p=\frac{|z|^{2\nu}-R^{2\nu}}{|z|^{2\nu}+R^{2\nu}}\quad,\quad q=\nu\arg(z)
\end{equation} 
Clearly we have $\Omega_{1}=0$, and for $\Omega_{2}$ one obtains
\begin{equation}
\Omega_{2}=\frac{1}{S^{2}R^{2}}\frac{2\pi^2}{3}\frac{\nu^{2}-1}
{\sin\frac{\pi}{|\nu|}}\quad,\quad |\nu|>1,
\end{equation}
and $\Omega_{2}=\frac{1}{S^{2}R^{2}}\frac{4\pi}{3}$ for $|\nu|=1$. 
In contrast to
the energy of the continuum model this quantity is not independent of $R$,
but scales with $\frac{1}{R^2}$. Again, for the semiclassical description
to be valid, $\sqrt{\Omega_{2}}$ should be small compared with the energy
$\langle{\cal H}_{3}\rangle=4\pi d$. This can be achieved for arbitrarily high
energies provided that the width of the instanton is sufficiently large.\\
Next we consider a solution consisting of two instantons separated by distance
$2l$,
\begin{equation}
w(z)=\left(\frac{z-l}{R}\right)^{\mu}\left(\frac{z+l}{R}\right)^{\nu}\,,
\end{equation}
and find
\begin{equation}
\Omega_{2}=\frac{1}{2S^2}\int d^{2}z
\left(\frac{|\frac{\partial w}{\partial z}|^{2}}
{\left(1+|w|^2\right)^2}\right)^2=\frac{1}{2S^{2}R^{2}}F(\mu,\nu;\frac{l}{R})
\end{equation}
with
\begin{equation}
F(\mu,\nu;a)=\int d^{2}z\left(
\frac{|z-a|^{\mu-1}|z+a|^{\nu-1}\left((\mu+\nu)z+(\mu-\nu)a\right)}
{1+|z-a|^{2\mu}|z+a|^{2\nu}}\right)^4\,, 
\end{equation}
Again, the square variance of the energy scales with $\frac{1}{R^2}$. 
Unfortunately, the quantity $F(\mu,\nu;a)$ cannot be expressed by elementary
functions; e.g. in the case $\mu=\nu$ one obtains after some algebra:
\begin{equation}
F(\mu,\mu;a)=a^{-8\mu-2}\int d^{2}x\frac{|\vec x|^{4\mu-4}|\vec x+\vec e|}
{\left(a^{-4\mu}+|\vec x|^{2\mu}\right)^4}
\end{equation}
with $\vec e$ being an arbitrary unit vector.\\\\
\underline{$(iv)$}\, Finally we consider an anisotropic two--dimensional
Heisenberg Hamiltonian:
\begin{equation}
{\cal H}_{4}=-J\sum_{m,n}\left[\hat{\vec S}_{m,n}\left(\hat{\vec S}_{m+1,n}
+\hat{\vec S}_{m,n+1}\right)-\lambda\hat S^{z}_{m,n}\hat S^{z}_{m,n}\right]
\label{defmod5}
\end{equation}
Choosing $\lambda>0$ and the lattice lying in the $x$-$y$-plane the classical
ground state is given by a parallel spin configuration in this easy plane.
Further we have: 
\begin{eqnarray}
 & & \langle{\cal H}_{4}^2\rangle-\langle{\cal H}_{4}\rangle^2=\nonumber\\
 & & (\hbar S)^{4}J^{2}\sum_{m,n}\Biggl[
\frac{1}{2S}\Biggl(\vec s_{m,n}\times\left(\vec s_{m-1,n}
+\vec s_{m+1,n}+\vec s_{m,n-1}+\vec s_{m,n+1}\right)\nonumber\\
 & & \hspace{2.8cm} 
-2\left(1-\frac{1}{2S}\right)\lambda\left(\vec s_{m,n}\cdot\vec e^{\,z}\right)
\left(\vec s_{m,n}\times\vec e^{\,z}\right)\Biggr)^2\nonumber\\
 & & +\frac{1}{8S^2}\Biggl(
\left(1-\vec s_{m,n}\vec s_{m-1,n}\right)^{2}
+\left(1-\vec s_{m,n}\vec s_{m+1,n}\right)^{2}
+\left(1-\vec s_{m,n}\vec s_{m,n-1}\right)^{2}
+\left(1-\vec s_{m,n}\vec s_{m,n+1}\right)^{2}\nonumber\\
 & & \hspace{3cm}+2\left(1-\frac{1}{2S}\right)\lambda^{2}
\left(\vec s_{m,n}\times\vec e^{\,z}\right)^{4}\Biggr)\Biggr]
\label{var5}
\end{eqnarray}
For sufficiently large $\lambda$ planar vortices are stable excitations 
(Gouvea {\it et al.} 1989). These objects are given in a continuum 
approximation by
\begin{equation}
p(\vec x,t)=0\quad,\quad q(\vec x,t)
=\nu\tan^{-1}\left(\frac{x_{2}}{x_{1}}\right)
\label{vor}
\end{equation}
with energy $E=\pi\nu^{2}\ln(\frac{L}{a})$, where we have chosen appropriate
units as before. The integer $\nu$ is called the vorticity, $L$ is the size of
the system and $a=1$ is the lattice spacing, which is a lower cutoff for
the integration. The energy diverges logarithmically with the system size, due
to the fact that the spin configuration is not parallel far away from the 
vortex center. In this sense, the planar vortex is not localized.\\
Inserting such a static planar solution into (\ref{var5}) obviously gives
$\Omega_{1}=0$ and in $\Omega_{2}$ a
contribution proportional to $\lambda^2$ that diverges quadratically with
growing $L$. This effect is completely analogous to the divergence in the 
model
(\ref{defmod3}) and has its reason in an unexact description of the quantum
ground state within our variational ansatz. So we renormalize $\Omega_{2}$ by
neglecting this artificial term, which is identical for any planar solution
and therefore in particular not sensitive for the vortex (\ref{vor}). 
The remaining expression for the quantum mechanical variance of energy is:
\begin{equation}
\Delta E=\frac{1}{2S}\sqrt{\frac{{\pi}}{6}}\nu^{2}\sqrt{1-\frac{1}{L^2}}
\label{vv}
\end{equation} 
In a classical easy--plane model of the above type a topological phase
transition, or Kosterlitz--Thouless--transition, occurs at a certain
temperature $T_{KT}$ (Kosterlitz and Thouless 1973, Kosterlitz 1974). 
Below this temperature vortices and 
antivortices (differing by the sign of the topological charge $\nu$) are 
bound in pairs, while they become mobile above $T_{KT}$.\\
Now we use our result (\ref{vv}) to estimate quantum corrections to
the value of the critical temperature $T_{KT}$. We assume that only 
vortices with $|\nu|=1$ are present in our system and follow some
rough approximations in the given references.\\
In the free energy $F=E-TS$ the quantum fluctuations will give an additional
contribution to the entropy $S$. For a single vortex the classical entropy is
$S=\ln(L^2)$, since the vortex can be placed anywhere in the system. In our
semiclassical description we have to take into account the uncertaincy of the 
energy:
\begin{equation}
S=\ln\left(L^{2}\left(1+\frac{\Delta E}{E}\right)\right)=2\ln L
+\frac{\Delta E}{E} +O\left(\frac{1}{S^2}\right)
\end{equation}  
The entropy term will dominate in the free energy above a temperature
\begin{equation}
T=\frac{\pi}{2}\frac{1}{1+\frac{\Delta E}{2\pi(\ln L)^{2}}}
=\frac{\pi}{2}\left(1-\frac{\Delta E}{2\pi(\ln L)^{2}}\right)
+O\left(\frac{1}{S^2}\right)\,,
\end{equation}
which reduces to the classical estimate $T_{KT}=\frac{\pi}{2}$ for 
$L\to\infty$. Thus, we conclude that in an infinite system there is no
quantum correction to the Kosterlitz--Thouless--temperature in first order
in $\frac{1}{S}$, but a logarithmic finite--size contribution arises in this 
order.

%%%%%%%%%%%%%%%%%%%%%%%%%%%%%%%%%%%%%%%%%%%%%%%%%%%%%%%%%%%%%%%%%%%%%%

\section{Conclusions}
In this work we investigate a large family of spin models in the
semiclassical regime with respect to quantum fluctuations.
Our main result is given by eqs.~(\ref{var1}),~(\ref{var2}) which provide
a physically relevant expression of quantum fluctuations of the energy
in low orders in $\frac{1}{S}$. In particular, the contribution in first 
order is purely due to the time--dependence of the spin configuration in
semiclssical description. For a static spin configuration the square variance
of the energy is of order $\frac{1}{S^{2}}$. The term of first order is
a natural property of a state vector with a non--trivial time--
dependence, which is consequently not an eigenstate of the Hamiltonian.
The terms of higher order can often be used as criterion of the validity
of the semiclassical approach.\\ 
These findings are valid
for a large variety of spin models in arbitrary dimension.
In the previous section we have illustrated this by several prominent
spin models.   
\\\\
{\bf Acknowledgement:} The authors are grateful to Holger Frahm 
for useful discussions.
 
%%%%%%%%%%%%%%%%%%%%%%%%%%%%%%%%%%%%%%%%%%%%%%%%%%%%%%%%%%%%%%%%%%%%%%

\newpage
\appendix{}
%\begin{appendix}

\section{Expectation values within coherent states}
\label{app1}

Let ${\cal P}[\hat{S^+},\hat{S^{-}},\hat{S^z}]$ be an arbitrary 
product of operator--valued spin components. With $U$ given in
eq.~(\ref{defU}) we have  
\begin{equation}
\langle S;\vartheta,\varphi|{\cal P}[\hat S^+,\hat S^{-},\hat S^z]
|S;\vartheta,\varphi\rangle=
\langle S|{\cal P}[U^+\hat S^+U,U^+\hat S^{-}U,U^+\hat S^zU]|S\rangle
\label{app1:prod}
\end{equation} 
It holds:
\begin{eqnarray}
U^+\hat S^+U=e^{\imath\varphi}
\left(\cos^{2}\left(\frac{\vartheta}{2}\right)\hat S^{+}-
\sin^{2}\left(\frac{\vartheta}{2}\right)\hat S^{-}+\sin\vartheta\hat S^z\right)
\label{app1:partU1}\\
U^+\hat S^zU=\left(-\frac{1}{2}\sin\vartheta\hat S^{+}
-\frac{1}{2}\sin\vartheta\hat S^{-}+\cos\vartheta\hat S^z\right)\,.
\label{app1:partU2}
\end{eqnarray}
Applying $\hat S^{+}$, $\hat S^{-}$ on states typed $|S-m\rangle$ gives 
a contribution of leading order $\sqrt S$, while $\hat{S^z}$ gives a 
factor of leading order $S$. By such arguments it is easily seen that 
the leading term in eq. (\ref{app1:prod}) is proportional to the product 
of the classical spin components given in $\vec s_{\vartheta,\varphi}$:
\begin{equation}
\lim_{S\to\infty\atop{\hbar S=S^{cl}={\rm const.}}}
\langle S;\vartheta,\varphi|{\cal P}[\hat S^+,\hat S^{-},\hat S^z]
|S;\vartheta,\varphi\rangle=
{\cal P}[S^{cl}s_{\vartheta,\varphi}^{+},
S^{cl}s_{\vartheta,\varphi}^{-},
S^{cl}s_{\vartheta,\varphi}^{z}]\,,
\end{equation}
where $(\hbar S)=S^{cl}={\rm const.}$ is the classical spin length. 
In the classical limit taken above all terms of higher order in 
$\frac{1}{S}$ (or equivalently in $\hbar$) drop. 
Of course, the details of these quantum contributions
depend on the structure of ${\cal P}[\hat S^+,\hat S^{-},\hat S^z]$,
i.e. the ordering of the spin operators.\\
Using eqs.~(\ref{app1:partU1}),(\ref{app1:partU2}) expectation values of type 
(\ref{app1:prod}) can be calculated easily. Here we give a list of
diagonal elements useful for the derivation of eq.~(\ref{var1}):
\begin{eqnarray}
\langle S;\vartheta,\varphi|\hat S^{z}\hat S^{z}\,
|S;\vartheta,\varphi\rangle & = & (\hbar S)^2\left(\cos^{2}\vartheta
+\frac{1}{2S}\sin^{2}\vartheta\right)\\
\langle S;\vartheta,\varphi|\hat S^{+}\hat S^{-}+\hat S^{-}\hat S^{+}\,
|S;\vartheta,\varphi\rangle
 & = & 2(\hbar S)^2\left(\sin^2\vartheta+\frac{1}{S}\left(
1+\cos^2\vartheta\right)\right)\\
\langle S;\vartheta,\varphi|\hat S^{+}\hat S^{z}+\hat S^{z}\hat S^{+}\,
|S;\vartheta,\varphi\rangle & = & 
2(\hbar S)^2\left(1-\frac{1}{2S}\right)
e^{\imath\varphi}\cos\vartheta\sin\vartheta\\
\langle S;\vartheta,\varphi|\hat S^{z}\hat S^{z}\hat S^{z}\,
|S;\vartheta,\varphi\rangle
 & = & (\hbar S)^3\left(\cos^3\vartheta+\left(\frac{3}{2S}
-\frac{1}{2S^2}\right)\cos\vartheta\sin^2\vartheta\right)\\
\langle S;\vartheta,\varphi|\hat S^{+}\hat S^{z}\hat S^{+}\,
|S;\vartheta,\varphi\rangle
 & = & (\hbar S)^3\left(1-\frac{1}{2S}\right)\left(1-\frac{1}{S}\right)
e^{\imath 2\varphi}\cos\vartheta\sin^2\vartheta\\
\langle S;\vartheta,\varphi|\hat S^{z}\hat S^{+}\hat S^{z}\,
|S;\vartheta,\varphi\rangle
 & = & (\hbar S)^3\left(1-\frac{1}{S}\right)e^{\imath\varphi}
\left(\cos^2\vartheta\sin\vartheta+\frac{1}{2S}\sin^3\vartheta\right)
\end{eqnarray}
\begin{eqnarray}
 &  & \langle S;\vartheta,\varphi|\hat S^{+}\hat S^{z}\hat S^{z}
+\hat S^{z}\hat S^{z}\hat S^{+}\,|S;\vartheta,\varphi\rangle
\nonumber\\
 & & \hspace{0.5cm}=2(\hbar S)^3e^{\imath\varphi}\left(\left(
1-\frac{1}{2S}\right)\left(1-\frac{1}{S}\right)
\cos^2\vartheta\sin\vartheta+\frac{1}{2S}\sin\vartheta\right) 
\end{eqnarray}
\begin{eqnarray}
 & & \langle S;\vartheta,\varphi|\hat S^{z}\hat S^{z}\hat S^{z}\hat S^{z}\,
|S;\vartheta,\varphi\rangle
\nonumber\\
 & & \hspace{0.5cm}=(\hbar S)^4\left(\cos^4\vartheta+\left(
\frac{3}{S}-\frac{5}{2S^2}+\frac{5}{8S^3}\right)\cos^2\vartheta\sin^2\vartheta
+\left(\frac{1}{2S^2}+\frac{1}{8S^3}\right)
\sin^2\vartheta\right) 
\end{eqnarray}
\begin{eqnarray}
 & & \langle S;\vartheta,\varphi|\hat S^{z}\hat S^{z}
(\hat S^{+}\hat S^{-}+\hat S^{-}\hat S^{+})
+(\hat S^{+}\hat S^{-}+\hat S^{-}\hat S^{+})\hat S^{z}\hat S^{z}\,
|S;\vartheta,\varphi\rangle\nonumber\\
 & & =2(\hbar S)^4\left(\left(2-\frac{6}{S}+\frac{5}{S^2}
-\frac{5}{4S^3}\right)\cos^2\vartheta\sin^2\vartheta+\frac{2}{S}\cos^2\vartheta
+\left(\frac{1}{S}+\frac{1}{4S^3}\right)\sin^2\vartheta\right) 
\end{eqnarray}
\begin{eqnarray}
 & & \langle S;\vartheta,\varphi|\hat S^{z}\hat S^{z}
(\hat S^{+}\hat S^{+}+\hat S^{-}\hat S^{-})
+(\hat S^{+}\hat S^{+}+\hat S^{-}\hat S^{-})\hat S^{z}\hat S^{z}\,
|S;\vartheta,\varphi\rangle\nonumber\\
 & & =2(\hbar S)^4\cos(2\varphi)\left(\left(2-
\frac{6}{S}+\frac{5}{S^2}-\frac{5}{4S^3}\right)
\cos^2\vartheta\sin^2\vartheta+\left(\frac{1}{S}-\frac{1}{4S^3}\right)
\sin^2\vartheta\right) 
\end{eqnarray}

\section{Coherent states and the classical limit of Spin systems}
\label{app2}

Let us consider a system of spins $\{\vec S_{i}\}_{i\in I}$. The dynamics
are given by a quantum Hamiltonian ${\cal H}[\{\hat{\vec S_{i}}\}_{i\in I}]$,
which is taken to be an arbitrary polynomial in the spin components.\\
We now evaluate the Heisenberg equation of motion for the $i$-th spin 
at a chosen time $t=t_0$ in terms of a tensor product of spin--coherent states
\begin{equation}
|\psi(t_0)\rangle=\bigotimes_{i\in I}\,|S_i;\vartheta_i,\varphi_i\rangle
\label{def2}
\end{equation}
and denote by $\langle\cdot\rangle$ an expectation value in the state
(\ref{def2}):
\begin{equation}
{\Big(\frac{d}{dt}\langle\hat{\vec S_{i}}\rangle\Big)}_{t=t_0}=
\frac{1}{\imath\hbar}{\Big(\langle[\hat{\vec S_{i}},\cal H]\rangle\Big)}_{t=t_0}
\end{equation} 
The right hand side of this equation can be evaluated straightforewardly
in terms of the variables $\vartheta_i$, $\varphi_i$, while for the
left hand side we need information about the time evolution of the 
wave function $|\psi(t)\rangle$.\\
The time evolution of the spin expectation values is
\begin{eqnarray}
\langle\psi(t)|\hat{\vec S_i}|\psi(t)\rangle=
\langle\psi(t_0)|\exp(+\frac{\imath}{\hbar}{\cal H}(t-t_0))
\hat{\vec S_i}\exp(-\frac{\imath}{\hbar}{\cal H}(t-t_0))|\psi(t_0)\rangle
\nonumber\\
={\langle\hat{\vec S_i}\rangle}_{t_0}
+\frac{\imath(t-t_0)}{\hbar}{\langle[{\cal H},\hat{\vec S_i}]\rangle}_{t_0}
+\frac{1}{2}\Big(\frac{\imath(t-t_0)}{\hbar}\Big)^2
{\langle[{\cal H},[{\cal H},\hat{\vec S_i}]]\rangle}_{t_0}+\cdots
\label{timeev}
\end{eqnarray}
The diagonal elements in the expansion (\ref{timeev}) are products 
of expressions of the form
(\ref{app1:prod}) and therefore reduce in the limit $S\to\infty$, 
$\hbar\to 0$, $\hbar S={\rm const}$ to the classical values. E.g. for
the first commutator we have
\begin{equation}
\lim_{S_{j}\to\infty\atop{\hbar S_{j}=S_{j}^{cl}={\rm const.}\atop{j\in I}}}
\!\!\!\frac{\imath}{\hbar}{\langle[{\cal H},\hat{\vec S_i}]\rangle}_{t_0}
=-S_{i}^{cl}\vec s_{i}
\times\frac{\partial{\cal H}[\{S_{j}^{cl}\vec s_j\}_{j\in I}]}
{\partial(S_{i}^{cl}\vec s_i)}
\end{equation} 
with $\vec s_i=(\sin\vartheta_i\cos\varphi_i,
\sin\vartheta_i\sin\varphi_i,\cos\vartheta_i)$. The right hand side of
this equation is nothing but the classical Poisson bracket 
$\{S_{i}^{cl}\vec s_i,{\cal H}\}$ occuring in the well--known
Landau--Lifshitz--equation, i.e. the equation of motion for a
classical spin system. Similar arguments hold for the higher
commutators in (\ref{timeev}), and we end up with
\begin{eqnarray}
\lim_{S_{j}\to\infty\atop{\hbar S_{j}=S_{j}^{cl}={\rm const.}\atop{j\in I}}}
\!\!\!\langle\psi(t)|\hat{\vec S_i}|\psi(t)\rangle=
S_{i}^{cl}\vec s_i+(t-t_0)\{S_{i}^{cl}\vec s_i,{\cal H}\}
+\frac{(t-t_0)^2}{2}\{\{S_{i}^{cl}\vec s_i,{\cal H}\},{\cal H}\}+\cdots
\end{eqnarray}
Thus, we have confirmed that the classical limit of the quantum mechanical
time evolution (\ref{timeev}) reproduces the motion of a classical
spin vector in the case of an arbitrary Hamiltonian.\\  
By similar considerations one can convince oneself that minimizing
the action
\begin{equation}
{\cal S}=\int_{t_1}^{t_2}dt
\langle\psi(t)|\imath\hbar\frac{d}{dt}-{\cal H}|\psi(t)\rangle
\label{action}
\end{equation}   
in the classical limit with respect to the functions 
$\vartheta_{i}(t)$, $\varphi_{i}(t)$ also leads to the
classical equations. This should be expected since the variation
principle concerning (\ref{action}) (with arbitrary wave function
$|\psi(t)\rangle$) is equivalent to the Heisenberg equation of
motion.

%\end{appendix}

%%%%%%%%%%%%%%%%%%%%%%%%%%%%%%%%%%%%%%%%%%%%%%%%%%%%%%%%%%%%%%%%%%%%%%

\end{document}